\documentclass[prl,aps,twocolumn,superscriptaddress]{revtex4-1}
\usepackage{graphicx,color}
\usepackage{amsthm}
\usepackage{amsfonts}
\usepackage{algorithmic}
\usepackage{enumerate}
\usepackage{latexsym}
\usepackage{amsmath}
\usepackage{amssymb}
\usepackage[colorlinks=true,citecolor=blue,linkcolor=blue]{hyperref}

\usepackage{dcolumn}
\usepackage{bm}

\usepackage[T1]{fontenc}
\usepackage{mathptmx}

\begin{document}
\title{Transport anisotropy and metal-insulator transition in striped Dirac fermion systems}

\author{Jingyao Meng}
\affiliation{Department of Physics, Beijing Normal University, Beijing 100875, China}
\author{Runyu Ma}
\affiliation{Department of Physics, Beijing Normal University, Beijing 100875, China}
\author{Lufeng Zhang}
\affiliation{School of Science, Beijing University of Posts and Telecommunications, Beijing 100876, China\\}
\author{Tianxing Ma}
\email{txma@bnu.edu.cn}
\affiliation{Department of Physics, Beijing Normal University, Beijing 100875, China}

\begin{abstract}
Using the determinant quantum Monte Carlo method, we investigate the metal-insulator transitions induced by the stripe of charge density in an interacting two-dimensional Dirac fermion system.
The stripe will introduce the transport anisotropy and insulating intermediate phase into the system, accompanied by the change of band structure and a peak of density of states around Fermi energy.
In the case of strong correlation, stripe exhibits competition with Coulomb repulsion through closing the energy gap and disrupting the magnetic order, and finally drives the system in Mott insulating phase back to the metallic state.
Our results may
provide a feasible way to modify transport properties by setting charge stripes in experiments.
\end{abstract}


\maketitle
\section{Introduction}

For the unique electronic spectrum\cite{Geim2007} and wonderful properties\cite{Ansell2015,LI2018845}, graphene has become one of the most promising Dirac fermion systems\cite{PhysRevLett.127.217001,PhysRevLett.125.176802}. With the emergence of a series of novel phenomena, the honeycomb lattice in correlated systems especially for Hubbard model is expected to reveal more complex physical mechanism. For example, the disorder in Anderson-Hubbard model is proven to induce a novel nonmagnetic insulating phase emerging at the quantum critical point\cite{PhysRevLett.120.116601}, the Hubbard-Holstein model including electron-phonon interaction identifies semimetal-to-insulator quantum critical points\cite{PhysRevB.103.235414}, and the Bose-Hubbard model is used to tune quantum phase transitions in helium-graphene system for its sensibility to the exact lattice structure\cite{PhysRevLett.126.107205}. In recent years, related researches have become increasingly richer, and numerous graphene-like lattices such as kagome\cite{C7TC02700A,PhysRevLett.128.096601} and decorated honeycomb lattice\cite{PhysRevLett.117.096405,PhysRevB.104.075104} have been discussed.

The stripe order composed by charge or spin inhomogeneities has received extensive attention in experimental and theoretical studies for its correlation with novel phenomena, such as symmetry breaking\cite{Huang2018}, superconductivity\cite{PhysRevLett.126.167001,10.21468/SciPostPhys.12.6.180}, topological phase\cite{PhysRevB.93.245308} and phase transitions\cite{PhysRevResearch.2.032071,PhysRevLett.98.046403,PhysRevA.93.011601}. Among them, the metal-insulator transition (MIT) has been an important and controversial issue.
For example, the MIT in hole-doped ferromagnets can be described as an ordering of the domain boundaries, which can be interpreted in terms of a 2D superstructure of orthogonal stripes\cite{PhysRevB.73.104453}.
In NdNiO$_{3}$, striped domains induced by heteroepitaxy change the surface morphology, and thus set the critical temperature of the first-order MIT\cite{Mattoni2016}.
Besides, the appearance of the stripe phase is accompanied by MIT under atomic scale investigation, indicating that the local conduction state is related to the charge ordering\cite{RENNER2005775}.

Recently, introducing charge stripes by setting the periodic potential energy has provided a powerful means to help us understand the physics of various lattices, especially for the graphene lattice.
In experiments, some organic materials are coated on graphite surface, and self-assemble into nanoscale stripes\cite{TEMIRYAZEV201930,Wastl2013,Wastl2014,Gallagher2016}. Although the size of stripe structure in graphene is rather small, its effect to induce the anisotropy is confirmed and highly valued.
Besides, numerical studies proposed that doping holes\cite{PhysRevB.103.155110} or modulating hopping\cite{Chen2021} could introduce the stripe order into graphene system,
and the stripe should cause interesting effects, such as its competition with the quantum anomalous Hall state in twisted bilayer graphene\cite{Chen2021}.
Actually, because of its ability to effectively adjust the physical properties of the system, charge stripes have been used to modify the electrical conductivity in graphene\cite{PhysRevLett.117.046603,Park2008}.
It has been also suggested that periodic potentials might lead to the generation of new Dirac points, and the changing energy band structure could be used to adjust the conductivity\cite{Lu2018}.
In experiments, highly pronounced resistance oscillations are found in the monolayer graphene with a laterally modulated potential profile\cite{PhysRevB.89.115421}, and the stripe built by the technique of dielectric patterning is proven to induce the transport anisotropy\cite{Li2021naturenano}.

In this paper, we study the stripe-induced metal-insulator transitions in the Hubbard model on a honeycomb lattice through the exact determinant quantum Monte Carlo (DQMC) method. The charge stripes are introduced by a periodically distributed chemical potential along the $y$-direction, and its strength is measured by $\Delta\mu$ as shown in Fig.~\ref{FigSigmaT}(h). A modulated potential is proven to effectively induce a stripe order\cite{xiao2022temperature}, and allows an exploration of general and fundamental issues\cite{PhysRevB.105.115116}. For the anisotropy of transport properties induced by stripe, we focus on the $x$-direction where there are more interesting physical phenomena.

Our data suggest that for the metallic system, the enhanced intensity of stripes induces two times phase transitions. That is, the increasing $\Delta\mu$ will first drive the semi-metal to a insulator, and then drive this insulator to a metallic phase. For a Mott insulator in the strongly correlated case, applying stripes closes the Mott gap and makes the system return to the metallic phase. We use the exact diagonalization method with no interactions to investigate the energy band structure influenced by stripes, and prove the existence of the insulating intermediate phase. Our results show that an increasing $\Delta\mu$ will lead to new Dirac points and a peak of density of states (DOS) which appears near the Fermi level\cite{Lu2018,PhysRevB.86.161108}. When $\Delta\mu$ continues to increase, the peak of DOS gradually disappears, and the system goes back to the metallic phase. The changing process of DOS coincides with the transition of transport properties, so we can label the insulating phase through the behavior of energy bands. Under a sufficiently large $\Delta\mu$, that is, a sufficiently strong stripe, the energy bands are separated from each other, and an energy gap appears at the Fermi level\cite{Park2008,cottam2019introduction,PhysRevB.101.035407}.
We summarize our results into a phase diagram as Fig.~\ref{Figphase}.


\section{Model and method}
\begin{figure*}[t]
\includegraphics[scale=0.41]{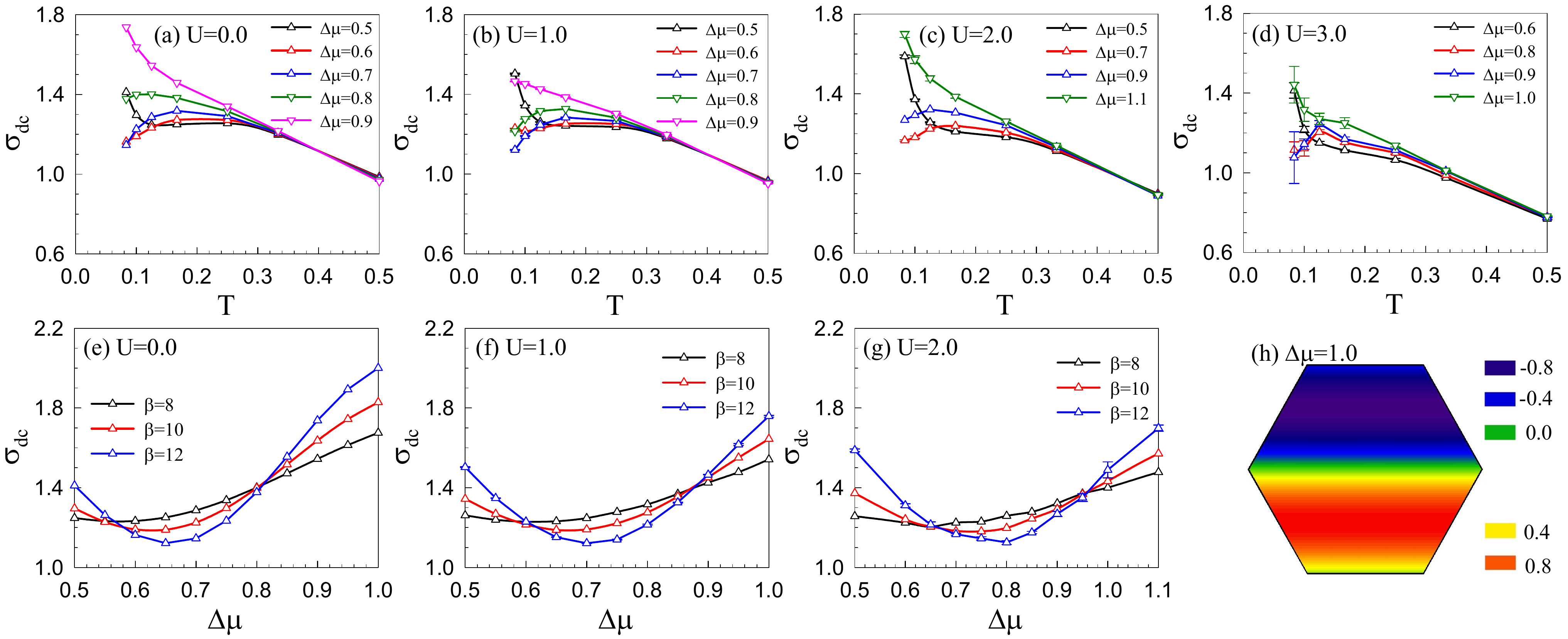}
\caption{\label{FigSigmaT} The transverse conductivity $\sigma_{dc}$ as a function of $T$ for several stripe strengths $\Delta\mu$ with interaction $U$ equal to: (a) 0.0, (b) 1.0, (c) 2.0 and (d) 3.0. At small $\Delta\mu$, $\sigma_{dc}$ decreases with increasing $T$, representing the metallic phase. At a pretty large $\Delta\mu$, $\sigma_{dc}$ decreases as $T$ decreases, indicating that metallicity is suppressed and MIT occurs. When $\Delta\mu$ is large enough, $\sigma_{dc}$ diverges again as the temperature decreased to the limit $T\rightarrow0$. Panels (e)$\sim$(g) show $\sigma_{dc}$($\Delta\mu$) curves for several interaction $U$ and inverse temperature $\beta$, whose intersections represent the critical points of phase transitions. Critical values of the first phase transition are respectively 0.58, 0.60, 0.65; critical values of the second phase transition are respectively 0.81, 0.88, 0.96. (h) The distribution of stripe along $y$-direction, taking $\Delta\mu$=1 as an example. The depth of color represents the value of chemical potential.}
\end{figure*}

The Hamiltonian of the interacting Hubbard model on a honeycomb lattice in the presence of charge stripes is defined as follows:
\begin{eqnarray}
\label{Hamiltonian}
\hat{H}&=&-t\sum_{\langle{\bf ij}\rangle\sigma}(\hat{c}_{{\bf i}\sigma}^\dagger\hat{c}_{{\bf j}\sigma}
+\hat{c}_{{\bf j}\sigma}^\dagger\hat{c}_{{\bf i}\sigma})+U\sum_{\bf j}(\hat{n}_{{\bf j}\uparrow}-\frac{1}{2})(\hat{n}_{{\bf j}\downarrow}-\frac{1}{2}) \nonumber\\
&&-\sum_{{\bf j}\sigma}\mu(\bf j)\hat{n}_{{\bf j}\sigma},
\end{eqnarray}
\begin{eqnarray}
\label{mu}
\mu({\bf j})=\Delta\mu\times\sin(2\pi(y({\bf j})-y_{0})/T_{y}).
\end{eqnarray}
In Eq.~\eqref{Hamiltonian}, $\hat{c}_{{\bf i}\sigma}^\dagger(\hat{c}_{{\bf i}\sigma})$ is the spin-$\sigma$ electron creation (annihilation) operator at site $\bf i$ and $\hat{n}_{{\bf i}\sigma}=\hat{c}_{{\bf i}\sigma}^\dagger\hat{c}_{{\bf i}\sigma}$ is the occupation number operator. Here, $t$ is the \textit{nearest-neighbor} (NN) hopping integral, and $t = 1$ sets the energy scale in the following. $U > 0$ is the onsite Coulomb repulsive interaction. $\mu({\bf j})$ is the chemical potential where $\Delta\mu$ describes the strength of the stripe. By setting the starting ordinate $y_{0}$ and the period length $T_{y}$ in the $y$-direction, $2\pi(y({\bf j})-y_{0})/T_{y}$ converts the ordinate $y({\bf j})$ into a stripe chemical potential distributed along the $y$ direction, and it forms a periodic distribution of the charge density. The schematic diagram is shown in Fig.~\ref{FigSigmaT} (h).

We adopt the DQMC method~\cite{PhysRevB.40.506} to study the phase transition in the model that is defined by Eq.~\eqref{Hamiltonian}, in which the Hamiltonian is mapped onto free fermions in 2D+1 dimensions that are coupled to space- and imaginary-time-dependent bosonic (Ising-like) fields. By using Monte Carlo sampling, we can carry out the integration over a relevant sample of field configurations, which are selected when the statistical errors are negligible enough.
The discretization mesh $\Delta\tau$ of the inverse temperature $\beta = 1/T$
should be small enough to ensure that the qualified Trotter errors are less than those that are associated with statistical sampling.
This approach enables us to compute static and dynamic observables at a specified temperature $T$.
Tuning $\mu({\bf i})\neq0$ means that the system is away from the half-filling, which breaks the particle-hole symmetry and will lead to a sign problem. However, the problem becomes less severe as we have $\sum_{{\bf j}}\mu({\bf j})=0$, and we are able to obtain accurate data at a large enough $\beta$ equal to $12$~\cite{PhysRevLett.120.116601}. We choose a $L = 6$ honeycomb lattice with periodic boundary conditions, for which the total number of sites is $N =2\times3\times L^2$.

\begin{figure}
\includegraphics[scale=0.41]{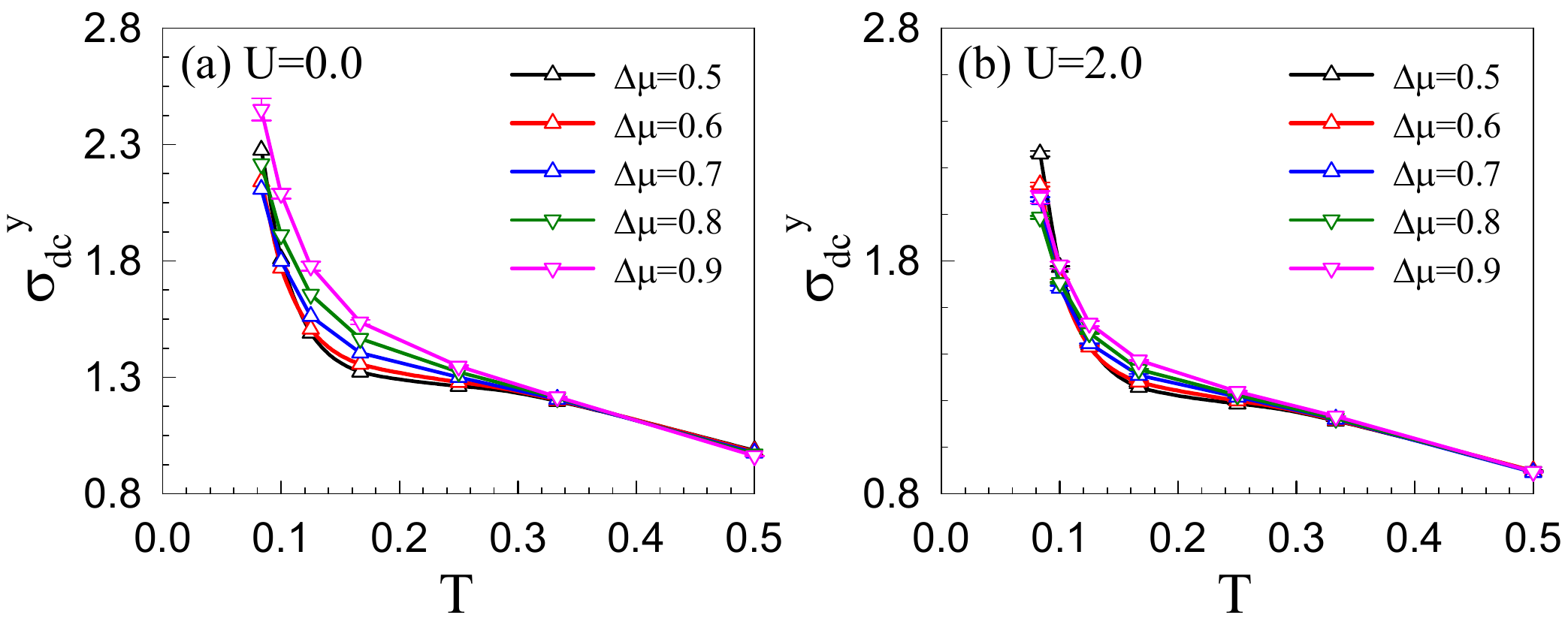}
\centering
\caption{\label{Figsigmay} The longitudinal conductivity $\sigma^{y}_{dc}$ computed as a function of temperature $T$ for various strengths of $U$ and $\Delta\mu$. The increasing stripe strength has little effect on $\sigma^{y}_{dc}$, the conductivity always diverges when the temperature tends to be 0, and the system is always metallic.
}
\end{figure}

The $T$-dependent $x$-direction dc conductivity is computed via a proxy of the momentum $\bf q$ and imaginary time $\tau$-dependent current-current correlation function (more details are in Appendix of Ref. \cite{PhysRevB.104.045138}):
\begin{eqnarray}
\label{conductivity}
\sigma_{\rm dc}(T)=\frac{\beta^{2}}{\pi}\Lambda_{xx}\left({\bf q}=0,\tau=\frac{\beta}{2}\right).
\end{eqnarray}
Here, $\Lambda_{xx}({\bf q},\tau)$ = $\langle \widehat{j_{x}}({\bf q},\tau)\widehat{j_{x}}(-{\bf q},0)\rangle$, and $\widehat{j_{x}}({\bf q},\tau)$ is the current operator in the $x$ direction. Similarly, the $y$-direction dc conductivity $\sigma_{\rm dc}^{y}(T)=\frac{\beta^{2}}{\pi}\Lambda_{yy}\left({\bf q}=0,\tau=\frac{\beta}{2}\right)$ describes the transport property in the $y$ direction as shown in Fig.~\ref{Figsigmay}.
This form, which avoids the analytic continuation of the QMC data, has been seen to provide satisfactory results for many studies~\cite{Scalettar1999,Mondaini2012}.

We also compute the staggered transverse antiferromagnetic(AFM) structure factor in the direction parallel to the lattice plane to study the AFM phase transition:
\begin{equation}
 S_{\rm AFM} = \frac{1}{N}\sum_{i,j}(-1)^{(i+j)} \left(S_i^{x}S_j^{x} + S_i^{y}S_j^{y}\right),
\end{equation}
where $S_i^{x}$ ($S_i^{y}$) is the $x$ ($y$)-component spin operator and the phase factor is $+1$($-1$) for sites $i$,$j$ that belong to the same (different) sublattices of the honeycomb structure.

The density of states at the Fermi level is defined as~\cite{PhysRevLett.75.312,Lederer2017}:
\begin{eqnarray}
\label{N0}
N(0)\simeq \beta \times G({\bf r} = 0,\tau = \beta/2).
\end{eqnarray}
The DOS is an important property to differentiate several physical mechanisms responsible for inducing the insulating phase, and $G$ in Eq.~\eqref{N0} is the imaginary-time dependent Green's function.
When $\Delta\mu$ becomes pretty large, the stripe will lead to the energy band deformation~\cite{PhysRevB.63.214513}, and $N(0)$ is not enough to describe the more complex condition around the Fermi energy.
Therefore, we need another method to study the change of DOS, especially at large $\Delta\mu$.

Using the exact diagonalization method, we calculated the energy bands of the lattice with no interaction, to help us understand the physical mechanism of the stripe-induced phenomena. We set a cell including 24 sites with different chemical potentials in a stripe period, and as $y({\bf j})$ increases, the site ${\bf j}$ is marked as $a_{1}$, $b_{1}$, $a_{2}$, $b_{2}$ ... $a_{12}$, $b_{12}$, respectively (more details are shown in Fig.~\ref{SMcell} of Appendix). The lattice is assumed to consist of a series of such cells distributed along the $x$-direction, so there is an approximate one-dimensional Hamiltonian defined as:
\begin{eqnarray}
\label{Hstripe}
\hat{H}&=&
-\sum_{{\bf i},j=1:12}t(\hat{c}_{{\bf i}a_{j}}^\dagger\hat{c}_{{\bf i}b_{j}}
+\hat{c}_{{\bf i}b_{j}}^\dagger\hat{c}_{{\bf i}a_{j}}
+\hat{c}_{{\bf i}b_{j}}^\dagger\hat{c}_{{\bf i}a_{j+1}}
+\hat{c}_{{\bf i}a_{j+1}}^\dagger\hat{c}_{{\bf i}b_{j}}) \nonumber\\
&&-\sum_{{\bf i},j=1:6}t(\hat{c}_{{\bf i}a_{2j}}^\dagger\hat{c}_{{\bf i+1}b_{2j}}
+\hat{c}_{{\bf i+1}b_{2j}}^\dagger\hat{c}_{{\bf i}a_{2j}} \nonumber\\
&&+\hat{c}_{{\bf i}a_{2j-1}}^\dagger\hat{c}_{{\bf i-1}b_{2j-1}}
+\hat{c}_{{\bf i-1}b_{2j-1}}^\dagger\hat{c}_{{\bf i}a_{2j-1}})
\end{eqnarray}
Here, ${\bf i}$ represents the $i_{th}$ cell, $a_{j}$($b_{j}$) is used to mark the $j_{th}$ site of sublattice $a$($b$) in a cell. For convenience, the hopping from $a_{1}$ to $b_{12}$ which represents periodic boundary condition on $y-$direction is included in $\hat{c}_{{\bf i}b_{j}}^\dagger\hat{c}_{{\bf i}a_{j+1}}
+\hat{c}_{{\bf i}a_{j+1}}^\dagger\hat{c}_{{\bf i}b_{j}}$. Through the second quantization, we get the matrix shown as Eq.~\eqref{H2nd} in which $\textbf{K}=e^{i\textbf{k}}$, and then get the energy bands by diagonalizing it numerically.

\begin{eqnarray}
\label{H2nd}
\footnotesize{
\begin{pmatrix}
\mu_{a_{1}} & 1+\frac{1}{\textbf{K}} & 0 & 0 & 0 & 0 & \ldots & 0 & 1 \\
1+\textbf{K} & \mu_{b_{1}} & 1 & 0 & 0 & 0 & \ldots & 0 & 0 \\
0 & 1 & \mu_{a_{2}} & 1+\textbf{K} & 0 & 0 & \ldots & 0 & 0 \\
0 & 0 & 1+\frac{1}{\textbf{K}} & \mu_{b_{2}} & 1 & 0 & \ldots & 0 & 0 \\
0 & 0 & 0 & 1 & \mu_{a_{3}} & 1+\frac{1}{\textbf{K}} & \ldots & 0 & 0 \\
0 & 0 & 0 & 0 & 1+\textbf{K} & \mu_{b_{3}} & \ldots & 0 & 0 \\
\vdots & \vdots & \vdots & \vdots & \vdots & \vdots & \ddots & \vdots & \vdots  \\
0 & 0 & 0 & 0 & 0 & 0 & \ldots & \mu_{a_{12}} & 1+\textbf{K}\\
1 & 0 & 0 & 0 & 0 & 0 & \ldots & 1+\frac{1}{\textbf{K}} & \mu_{b_{12}}\\
\end{pmatrix}}
\end{eqnarray}

\section{Results and Discussion}

\begin{figure}[t]
\includegraphics[scale=0.41]{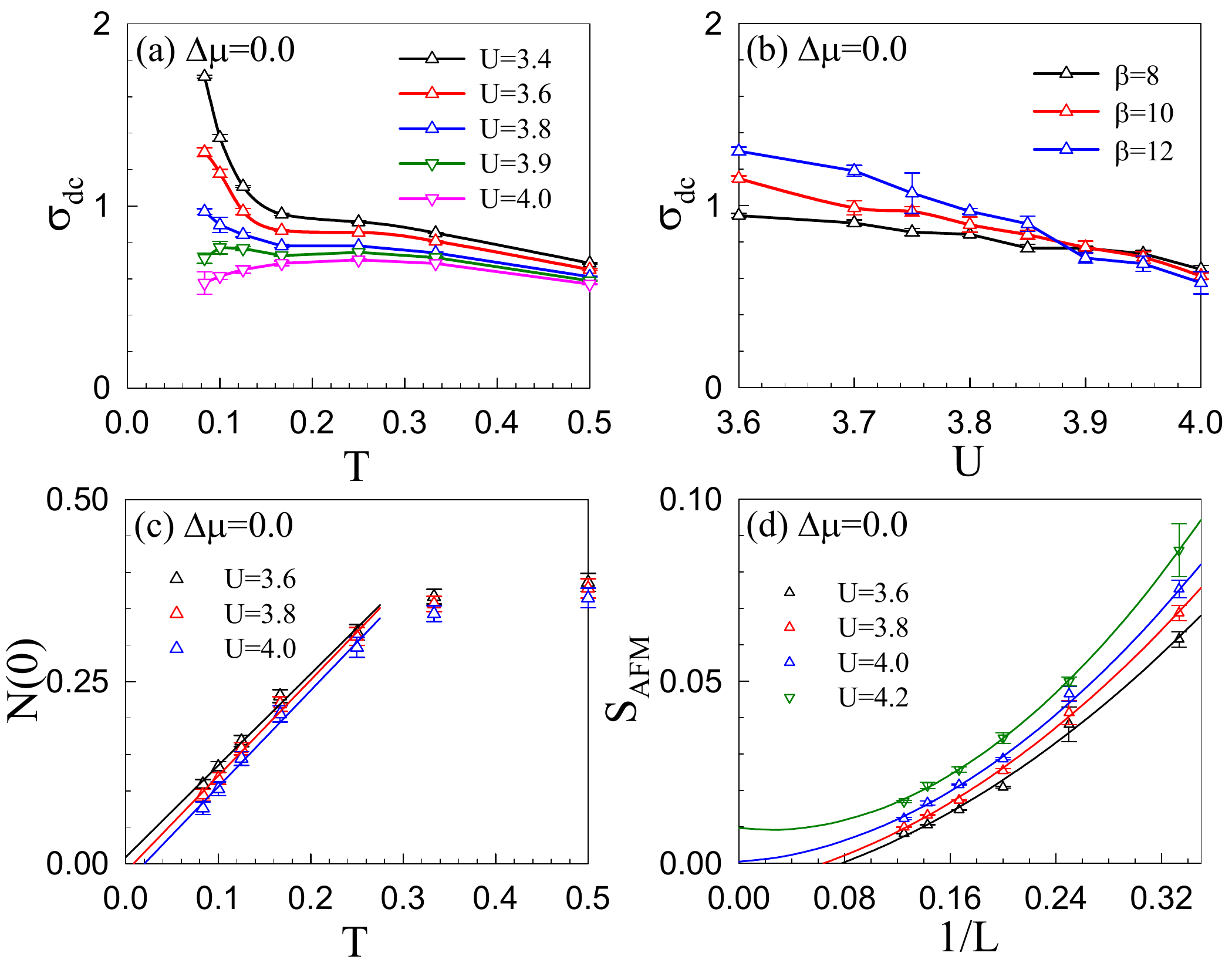}
\caption{\label{FigUc}
(a) Conductivity $\sigma_{dc}$ as a function of temperature $T$ for several interaction $U$ at $\Delta\mu$=0. $U$ inhibits metallicity and induces MIT, and conductivity increases with increasing temperature. (b) $\sigma_{dc}$ as a function of $U$ for various inverse temperature $\beta$. The curves of $\sigma_{dc}(T)$ intersect around $U\approx3.89$, the function relationship between $\sigma_{dc}$ and $T$ is different on two sides of the intersection, representing the critical value for inducing MIT. (c) DOS at the Fermi energy $N(0)$ as a function of temperature $T$. At a sufficiently large $U$, $N(0)$ tends to be zero when $T\rightarrow0$, which suggests an opened Mott gap. We use points at low temperature for a linear fit. (d) Staggered transverse AFM structure factor $S_{AFM}$ as a function of lattice size $L$ for various $U$. As the interaction increases, $S_{AFM}$ is increased at each $L$. As the curve intercept gradually increases from zero to positive, the system reaches the AFM phase. The critical $U$ is approximately 3.9. For finite-size scaling studies, quadratic fit is proved to be suitable\cite{PhysRevB.37.2380,PhysRevX.3.031010}.}
\end{figure}

\begin{figure}[t]
\includegraphics[scale=0.41]{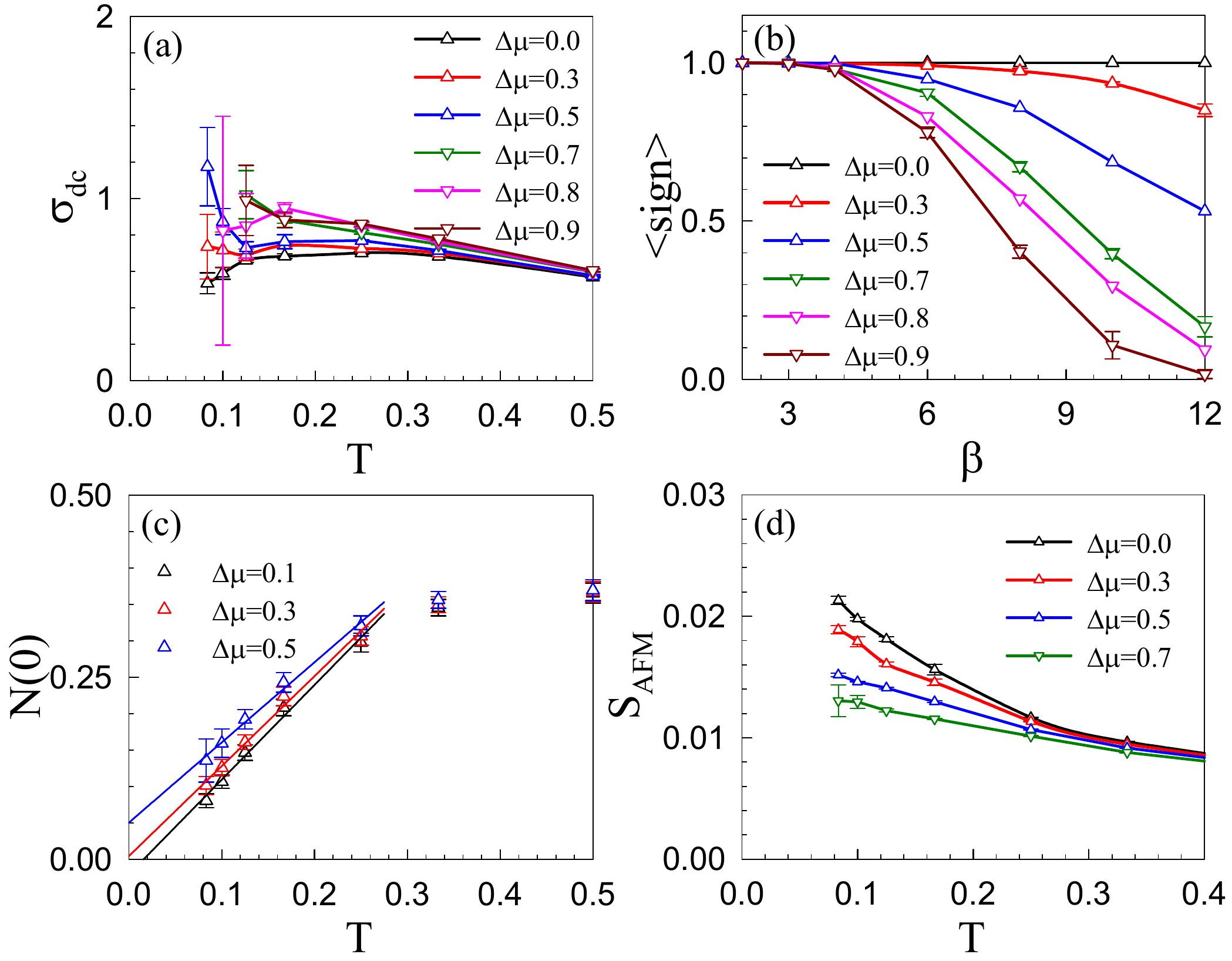}
\caption{\label{FigUdmu} For interaction $U=4.0$: (a) Conductivity $\sigma_{dc}$ as a function of temperature $T$ for several stripe strength $\Delta\mu$. When $\Delta\mu$ increases from 0.0 to 0.7, it enhances metallicity and induces insulating-metallic phase transition. At $\Delta\mu=0.8$, the system is a stripe-induced insulator. Stripe continues to enhance, $d\sigma_{dc}/dT$ becomes positive, and the error bar is large under sufficiently strong $U$ and $\Delta\mu$. (b) The sign problem $\langle$$sign\rangle$ as a function of the inverse temperature $\beta$. The larger the $\Delta\mu$, the faster the $\langle$$sign\rangle$ decreases with $\beta$, so our calculations are limited under large $U$ and large $\Delta\mu$. (c) DOS at the Fermi energy $N(0)$ as a function of temperature $T$. As $\Delta\mu$ increases, the $N(0)$ curve tends to be an infinite value at $T\rightarrow0$. (d) Staggered transverse AFM structure factor $S_{AFM}$ as a function of temperature $T$ at different stripe strength $\Delta\mu$. As $\Delta\mu$ increases, the divergent $S_{AFM}$ at low temperature is suppressed.}
\end{figure}

Starting from the metallic graphene system under weak interaction\cite{PhysRevB.104.045138}, our results show that the application of stripe will introduce two times phase transitions into the system, as shown in Fig.~\ref{FigSigmaT}.
In panels (a)$\sim$(d), the conductivity $\sigma_{dc}$ (\textit{"conductivity"} means \textit{"lateral conductivity"} unless otherwise specified) is a function of temperature $T$ for several $U$ and $\Delta\mu$. While the conductivity decreases at lower temperatures in the (semi-) metallic phase with sufficient small interaction, the effect of an increase of $\Delta\mu$ is unequivocal. The stripe first leads to a suppression of metallicity accompanied by MIT, displaying a downturn of $\sigma_{dc}$ at small $T$. When $U=0$, increasing $\Delta\mu$ from 0.5 to 0.7 introduces insulating phase into the system. However, when stripe continues to increase, an opposite phenomenon emerges: $d\sigma_{dc}/dT$ changes from negative to positive. As $\sigma_{dc}$ decreases with increasing $T$ at low temperatures, the system re-enters metallic phase. We use MIMT(metal-insulator-metal transition) to define this process.
In panel (b) to (d), we calculated the situation under $U=1\sim3$, and it can be indicated that MIMT is effective whether there is an interaction or not.

\begin{figure*}[t]
\includegraphics[scale=0.35]{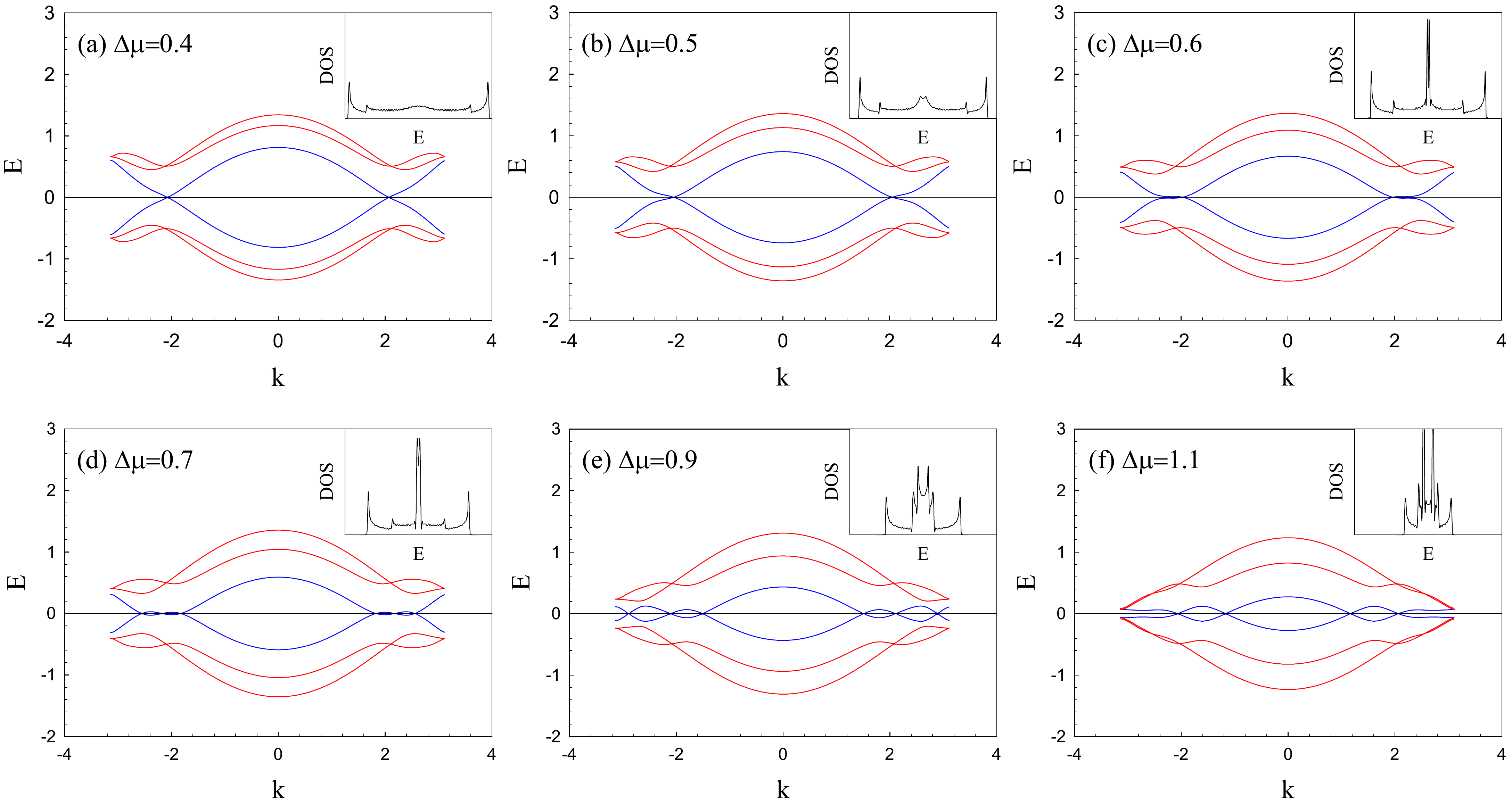}
\caption{\label{Figband} Band structure and DOS distribution near the $E_{0}$ without interaction. When stripe is applied, the energy band near the Fermi surface is deformed. As the $\Delta\mu$ increases, the two blue bands approach the $E_{0}$ (as panels (b) and (c)) and generate new crossings (as panel (d)), accompanied by a peak at $E=0$ for DOS. The peak splits and disappears as $\Delta\mu$ increases to 1.1. In (c) and (d), the DOS peak is confined to a small range, the magnitude order of peak width $\Delta$$E$ is about $10^{-2}$.}
\end{figure*}

A more evident display of the critical stripe strength for MIMT is obtained in panels (e)$\sim$(g), where $\sigma_{dc}$ is a function of $\Delta\mu$ for various inverse temperatures $\beta$. Since the phase transition implies a change in the function relationship between $\sigma_{dc}$ and $T$, the intersections of the $\sigma_{dc}(\Delta\mu)$ curves can be used to estimate the critical $\Delta\mu$. When the curve with lower temperature (like blue curve for $\beta$=12) is higher than the curve with large temperature (like black curve for $\beta$=8), the system is metallic. And the region where the blue curve is lower than the black one indicates the presence of the intermediate insulating phase.

Moreover, in the direction parallel to the stripe, the different chemical potential distributions lead to different conditions of electron hopping, so we need to calculate the behavior of longitudinal conductivity $\sigma^{y}_{dc}$.
As shown in Fig.~\ref{Figsigmay}, when $\Delta\mu$ increases from 0.5 to 0.9, $\sigma^{y}_{dc}$ in $y-$direction hardly changes whether or not there is an interaction, which is very different from the MIMT on the $x-$direction. $\sigma^{y}_{dc}$ always increases with decreasing temperature, meaning that the system is always a metal. By contrast, our results suggest that the application of stripe contributes to the formation of transport anisotropy, which is also reported in Ref. \cite{PhysRevB.63.214513,Park2008,Li2021naturenano}.

Thus we focus on the transport properties on $x-$direction. For the graphene system in the Hubbard model, Mott insulating phase under strong correlation is a crucial issue, and its interplay with stripe order needs further study\cite{PhysRevB.103.155110}.
First, the Mott insulating phase caused by the interaction $U$ are determined in Fig.~\ref{FigUc}. Panel (a) plots $\sigma_{dc}$ as a function of $T$, where $\sigma_{dc}$ changes little with $T$ as $U$ increases to $3.8\sim3.9$, and the system is close to the critical point of MIT. In Fig.~\ref{FigUc}(b), the conductivity always decreases as the interaction increases, and $\sigma_{dc}$ at small $T$ decreases faster. Around the critical value $U_{c}\approx3.89$, $d\sigma_{dc}/dT$ changes from positive to negative, and the interaction drives the metal to Mott insulating phase.
In addition to the MIT, we investigated the change in DOS to determine the behavior of the energy gap. $N(0)$ is the function of $T$ in panel (c), and the value of $N(0)$ at $T\rightarrow0$ curve decreases gradually as $U$ increases. Under a sufficiently large $U$, $N(0)\rightarrow0$ when $T\rightarrow0$, which means that there is no electron distribution near the Fermi energy $E_{0}$, and the Mott energy gap is opened\cite{PhysRevLett.117.146601}. We linearly fitted the data at low temperatures, and the critical point determined by $N(0)$ is around 3.8.
Similarly, we also extrapolate the data to the thermodynamic limit in panel (d), and determined that symmetry breaking and antiferromagnetic phases appear around $U\approx4.0$: In the finite-size scaling study of the AFM spin structure factor $S_{AFM}$ in Fig.~\ref{FigUc}(c), the value of $S_{AFM}$ at $L\rightarrow\infty$ is 0 when $U=$3.6 and 3.8. When $U\geq4.2$, $S_{AFM}$ tends to be a finite value, which proves that the antiferromagnetism at this time is indeed a long-range order.
Besides, although these $U-$driven MIT points determined by different methods are not coincided, the critical interaction $U_{c}$ can be determined around $3.9$.
Therefore, we can conclude that the Mott gap is opened by the strong Coulomb repulsion at $U_{c}$, accompanied by a MIT and a magnetic phase transition. The critical value $U_{c}\approx3.9$ is consistent with the conclusions of the studies\cite{Sorella2012Uc3.9, PhysRevLett.120.116601}.

Next, we discuss the competition between stripe and Mott insulating phase as shown in Fig~\ref{FigUdmu}, which involves conductivity, magnetic order and band structure.
In Fig~\ref{FigUdmu}(a), the conductivity $\sigma_{dc}$ increases with the temperature $T$ at $\Delta\mu$=0, and the system is in the Mott insulating phase under strong interaction $U=4$\cite{PhysRevB.104.045138,PhysRevLett.120.116601}. When the periodic chemical potential is applied and its oscillation amplitude becomes stronger, the metallicity of the system is promoted by $\Delta\mu$. As $\Delta\mu$ reaches 0.3, $d\sigma_{dc}/dT$ tends to be zero, representing that the system is at the critical point of the Mott insulator-metal transition. As $\Delta\mu$ increases to 0.5, the sysrem becomes a distinct $\Delta\mu$-dominant metal. When $\Delta\mu > 0.5$, the system will undergo MIMT as described in Fig.~\ref{FigSigmaT}.
It is worth noting that, the determination of these two transitions at $U=4$ is slightly more problematic, for the large sign problem accompanied by the large error bar. As shown in panel (b), $\langle$$sign\rangle$ at large $\Delta\mu$ drops rapidly to 0 with decreasing temperature, so we can only qualitatively conclude that there exists the MIMT. However, although the large error under pretty strong $U$ will limit the in-depth analysis, the sign problem at $\Delta\mu$=0.3 is not serious and the conclusion that the stripe could suppress the Mott insulating phase is reliable.
By fitting $N(0)$ at low temperature as a linear function of $T$, we find that as $\Delta\mu$ increases, $N(0)_{T\rightarrow0}$ increases gradually, and at the critical $\Delta\mu$ =0.3 it becomes positive. At this time, DOS near the Fermi surface is not 0, the electron distribution reappears at $E_{0}$, and the energy gap is closed.
In addition, stripe will suppress the magnetic order and eliminate the AFM phase. For an antiferromagnet, $S_{AFM}$ is expected to diverge as $T\rightarrow0$\cite{APLmtx2010, li2022metalinsulator}. In panel (d), $\Delta\mu$ exhibits an obvious effect to inhibit $S_{AFM}$, especially for the low temperature. Although the existence of period length $T_{y}$ limits us to do a finite-size scaling study, the inhibition on AFM by $\Delta\mu$ is obvious. Overall, stripe will close the Mott gap and eliminate the magnetic order, inducing the transition from Mott insulating phase to metal.

So far we have discussed how the stripe competes with the interaction through dc conductivity.
Another electronic property to characteristic the system state is the DOS.
We therefore simplify the lattice model to a 2D Hamiltonian as shown in Eq.~\eqref{Hstripe},
resolve the quadratic quantization matrix and get the band distribution of the system as shown in the appendix.
We focus on the four dominating energy bands around the $E_{0}$ in Fig.~\ref{Figband},
and study the distribution of DOS near $E_{0}$ based on the band structure which are closest to the Fermi surface, indicated by the two lines in blue color.
As shown in Fig.~\ref{Figband}, with increasing $\Delta\mu$, the width of these two bands reduce. Around two Dirac points, they gradually approach and lead to new crossings(in other words, new Dirac points) on the Fermi surface as shown in panel (c). For periodic potentials on the 2D lattice, Ref. \cite{PhysRevB.86.161108,Lu2018} exhibits a similar effect, which provides support for our conclusion. Subsequently, the two bands indicated in red color gradually approach the Fermi level and ``touch'' the bands indicated in blue when $\Delta\mu$ is pretty large, such as $\Delta\mu =$ 1.1. Our results in Fig.~\ref{SMband} suggest that a strong enough stripe may be accompanied by an energy gap as described in previous studies\cite{PhysRevB.101.035407,PhysRevLett.102.047001}. Here we focus on the MIMT in the development of stripe. It is worth noting that the change of energy band affects the distribution of DOS, whose change trend is consistent with the transport properties.

For panel (a) to panel (b), the system is metallic under weak stripe. Although the change in the band structure leads to an increase in DOS at $E_{0}$, the peak of DOS around Fermi level does not appear until $\Delta\mu$ increases to 0.6. In particular, the metal-insulator transition also occurs at this value as described in Fig.~\ref{FigSigmaT}(e). In the insulating phase, like $\Delta\mu$=0.7 in panel (d), the sharp peak at $E_{0}$ appears. However, with further enhancement of the stripe and compression of the bandwidth, beyond the crossings, two energy bands indicated in blue color are gradually distanced. Although the DOS at $E_{0}$ still maintains a large value, the ``diffusion'' of the DOS distribution still leads to the disappearance of the peak, with the return of the system to the metallic phase. We take lattice at $\Delta\mu$=0.9 and 1.1 as an example, whose conductivity behavior is in Fig.~\ref{FigSigmaT}(e) and energy band behavior is in Fig.~\ref{Figband}(e)$\sim$(f).

\begin{figure}[t]
\includegraphics[scale=0.55]{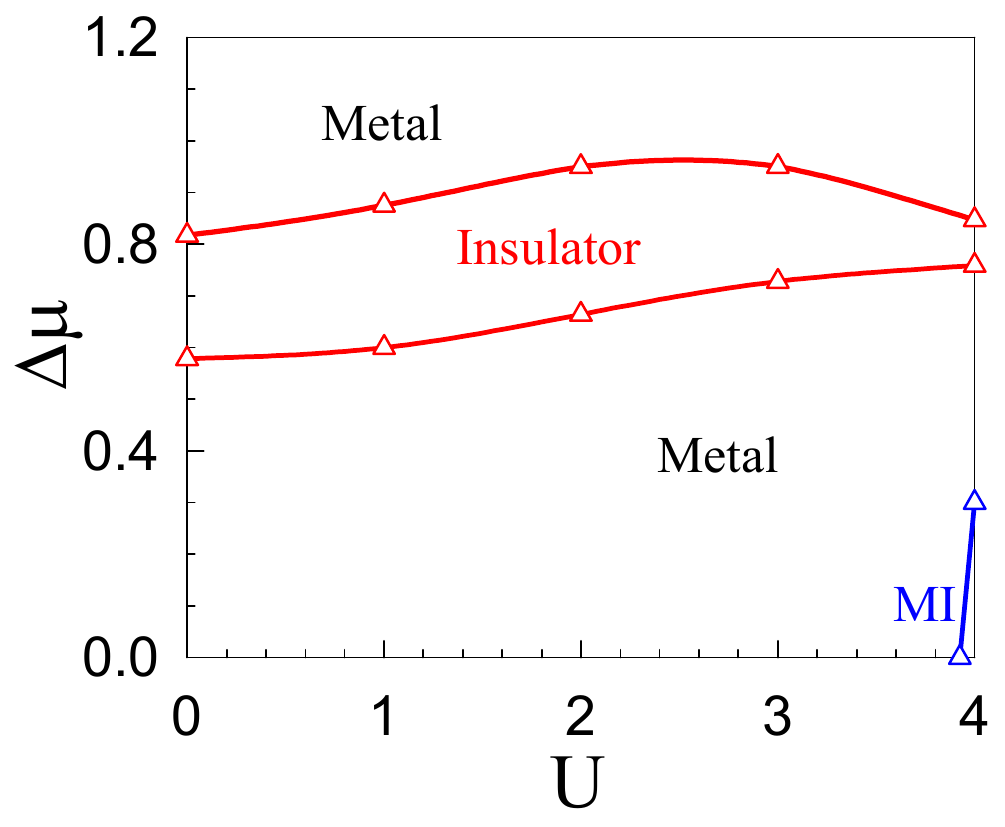}
\caption{\label{Figphase} Critical stripe strength $\Delta\mu_{c}$ under different interaction $U$. The phase diagram is divided into four regions: metal under weak U and $\Delta\mu$, Mott insulator under strong $U$ weak $\Delta\mu$, metal phase under strong $\Delta\mu$, and intermediate state which is a $\Delta\mu$-induced insulating phase.}
\end{figure}

With results of energy bands supporting our previous conclusions, we summarize our results as a phase diagram shown in Fig.~\ref{Figphase}. The $\Delta\mu$-induced insulating phase is wrapped by two red curves, and with the increase of $U$, the two red curves have a tendency to close, indicating that the intermediate state may be suppressed by strong interaction, and the system is a $U-$dominated Mott insulating phase.
Due to the sign problem and error bar, the calculation of competition between Mott insulator and stripe is limited in a small range, and labeled by the blue curve. On the right side of this curve, conductivity increases with temperature, and the system has AFM order and energy gap. As $U$ increases, the blue curve rises rapidly, which also proves the instability of other phases under strong correlation.

\section{Summary}

Using DQMC simulations, we investigated the effect of stripe on the transport properties in the Hubbard model. Through a periodically distributed chemical potential along the $y-$direction, we introduce charge stripes into the 2D honeycomb lattice, and define its intensity as $\Delta\mu$.
The change in lateral conductivity $\sigma_{dc}$ as a function of temperature $T$ indicates that stripe will induce an insulating intermediate state and two phase transitions. Through secondly quantizing the simplified non-interacting model, we use the behavior of DOS to verify the existence of mesophase. We speculate that the reason for the stripe-induced intermediate phase may be the change in band structure. However, the stripe causes different potential distributions on $x-$ and $y-$direction, and the longitudinal conductivity is hardly affected by $\Delta\mu$. Thus, the system will exhibit transport anisotropy.

Stripe also showes competition with interaction. Through conductivity, DOS at the Fermi level, and antiferromagnetic spin structure factor, we demonstrate that for a strongly correlated Mott insulator, a sufficiently strong stripe will close the Mott gap, disrupt the magnetic order, and ultimately drive the system as a metal. We summarize our results in the phase diagram including metal, Mott insulator and stripe-induced insulating phase as shown in Fig~\ref{Figphase}, providing a discussion and reference for modifying transport properties by setting charge stripes.

\noindent
\underline{\it Acknowledgments} ---
We thank Rubem Mondaini for many helpful discussions.
This work was supported by the NSFC (Nos. 11974049 and 11734002) and NSAF U1930402. The numerical simulations were performed at the HSCC of Beijing Normal University and on Tianhe-2JK in the Beijing Computational Science Research Center.

\appendix

\setcounter{equation}{0}
\setcounter{figure}{0}
\renewcommand{\theequation}{A\arabic{equation}}
\renewcommand{\thefigure}{A\arabic{figure}}
\renewcommand{\thesubsection}{A\arabic{subsection}}

\section{The DC conductivity}
\label{app:DCcon}

\begin{figure}[t]
\includegraphics[scale=0.38]{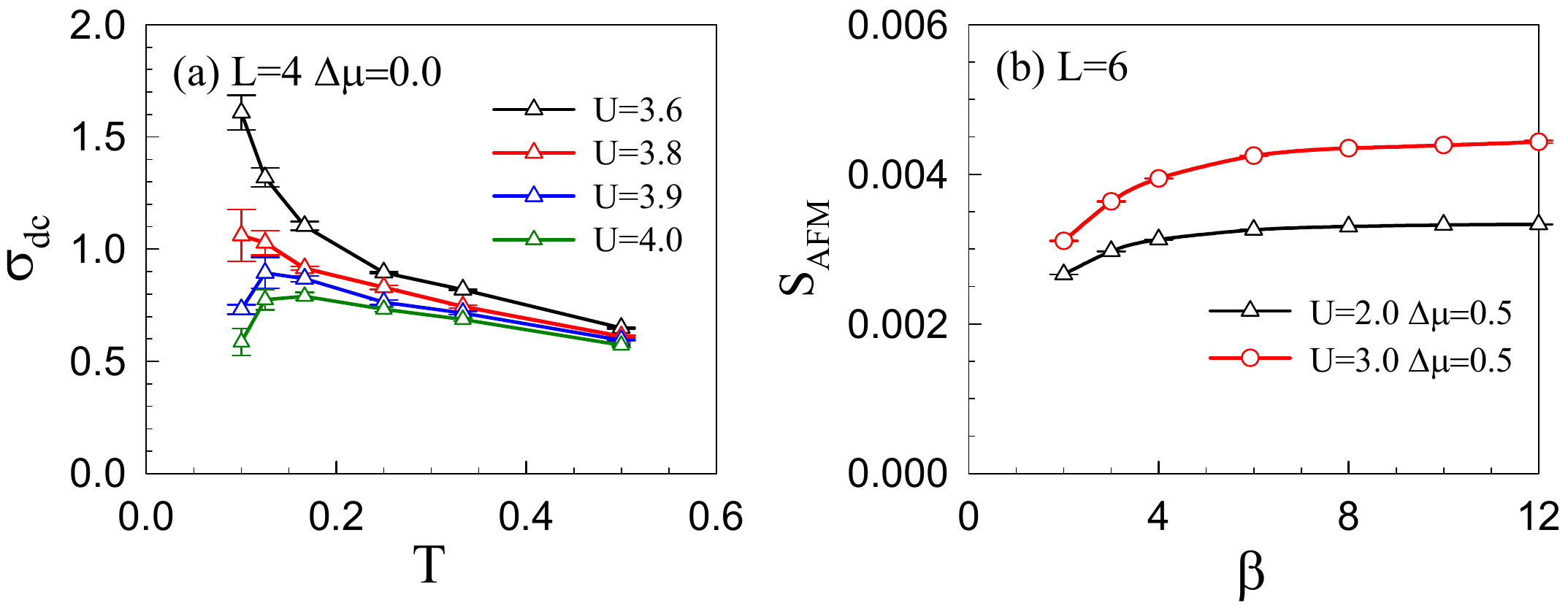}
\centering
\caption{\label{SMTL} (a) The conductivity $\sigma_{\rm dc}$ is shown as a function of temperature $T$ for various interaction $U$ with lattice size $L=4$.
(b) Staggered transverse AFM structure factor $S_{AFM}$ as a function of $\beta$. We choose a large enough $T=1/12$ in our calculation.}
\end{figure}

To understand the influence of the system's finiteness on the physical results we have presented in the main text, we now check the behavior of $\sigma_{dc}$ with different lattice size $L$. We report in Fig.~\ref{SMTL}(a) the conductivity $\sigma_{dc}$ as a function of temperature $T$ for $L=4$. When we compare Fig.~\ref{SMTL}(a) and Fig.~\ref{FigUc}(a), we can find that while different lattice sizes yield different values for the conductivity, the interaction still induces a Mott insulating phase at the critical value around $3.8\sim3.9$. Besides, the system is always a metal when $U=3.6\sim3.8$ but an insulator when $U=3.9\sim4.0$.

To test that at $\beta$=12 we are already assessing physics close to the ground state, we show in Fig(b) the dependence of $S_{AFM}$ with the inverse temperature: saturation is readily observed for values $\beta\gtrsim10$.

\begin{figure}[t]
\includegraphics[scale=0.3]{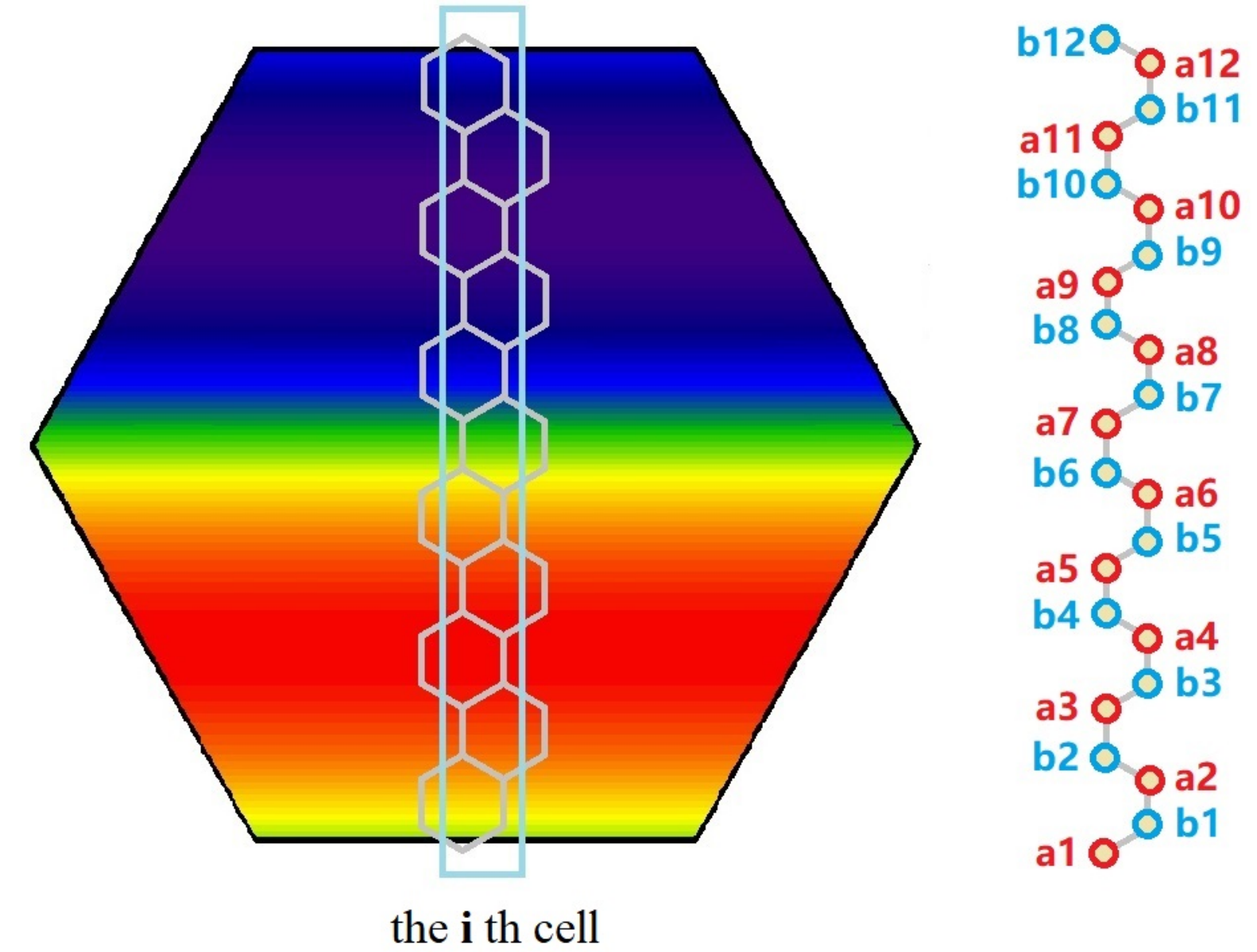}
\centering
\caption{\label{SMcell} The cell of simplified Hamiltonian including 24 sites with different chemical potentials. Sublattices are labeled by red and blue colors.}
\end{figure}

\section{The simplified Hamiltonian}\label{app:realz}
\begin{figure}
\includegraphics[scale=0.28]{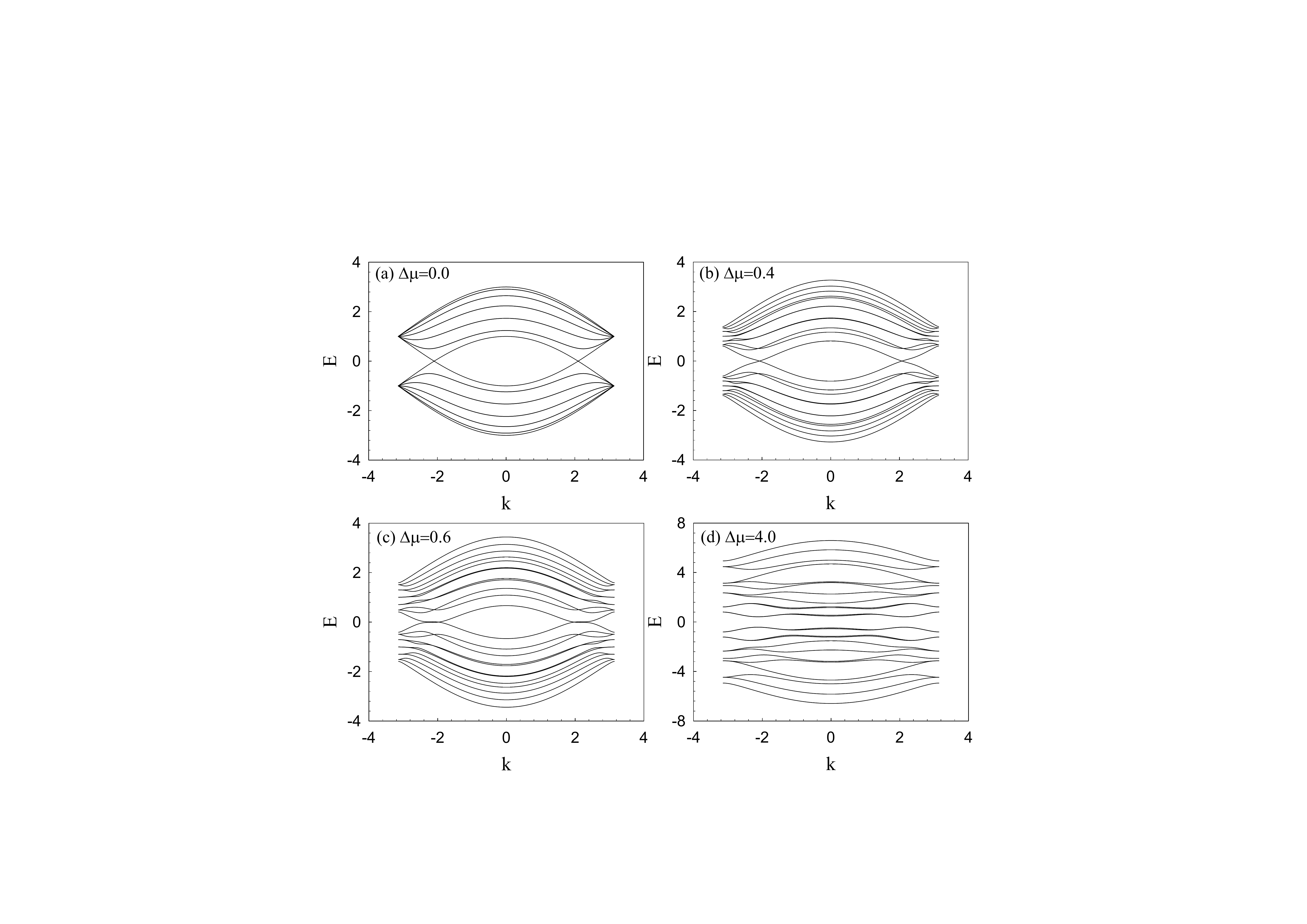}
\centering
\caption{\label{SMband} Energy bands $E(k)$ at: (a) $\Delta\mu=0.0$. Some bands are degenerate and the DOS near Fermi energy vanishes linearly. (b) $\Delta\mu=0.4$. The energy levels begin to deform, and the system is metallic. (c) $\Delta\mu=0.6$. The system is at the critical point of phase transition. When $\Delta\mu$ continues to increase, new crossings will appear on the Fermi energy as shown in Fig.~\ref{Figband}(d). (d) $\Delta\mu=4.0$. The stripe is very strong and energy bands are separated from each other.
}
\end{figure}

We set the cell as shown in Fig.~\ref{SMcell}, which contains a complete cycle of chemical potential distribution with 24 sites. For sites on the $2\times3\times6^{2}$ graphene lattice, this cell includes all possible potentials under a determined $\Delta\mu$.
In the simplified model, we assume that these cells are aligned along the $x-$direction, with only one cell in the $y-$direction, and periodic boundary conditions are effective on both directions. Therefore, the system is reduced to a one-dimensional Hamiltonian as shown in Eq.~\eqref{Hstripe}, where ${\bf i}$ and $a_{j}$($b_{j}$) are respectively used to represent the ${\bf i}$ th cell and the $j$ th a(b) site in this cell. In Eq.~\eqref{Hstripe}, the first term is the transition within the cell, such as $a_{1}\rightarrow b_{1}$, $b_{1}\rightarrow a_{2}$, $b_{12}\rightarrow a_{1}$. The second term is the transition between cells, such as $b_{1}$ in ${\bf(i-1)}$ th cell $\rightarrow a_{1}$ in ${\bf i}$ th cell, $a_{2}$ in ${\bf i}$ th cell $\rightarrow b_{2}$ in ${\bf (i+1)}$ th cell.

Through the second quantization, we obtain the Hamiltonian matrix as shown in Eq.~\eqref{H2nd} and $E(k)$ at no-interacting case. We show all energy bands in Fig.~\ref{SMband}, and focus on the situation near the Fermi energy. Although the interaction in our calculation is not zero, the value range of interaction is not particularly large ($U\leq4$), so we think this method can still be used for reference.
Panel (d) shows the separation of bands at large $\Delta\mu$, indicating that although the spatial period of the potential is much larger than the distance between two nearest sites, a strong enough stripe will also induce an energy gap\cite{PhysRevLett.83.5098,Park2008,cottam2019introduction}. It is worth noting that the $y-$direction periodic boundary condition destroys the possible edge states in the 2D graphene ribbons, which only has periodic boundary condition in one direction.

\bibliography{References}

\end{document}